\documentclass[preprint,5p,times,twocolumn]{elsarticle}
\usepackage{tabularx}
\usepackage[utf8]{inputenc}
\usepackage{amsmath}
\usepackage{soul, color}

\usepackage{wrapfig}
\usepackage{graphicx}
\usepackage{glossaries}
\usepackage{glossary-mcols}
\glsdisablehyper
\makeglossaries
\newacronym{CT}{CT}{Computed Tomography}
\newacronym{SVM}{SVM}{Support Vector Machine}
\newacronym{CHS}{CHS}{Cardiovascular Health Study}
\newacronym{MIT}{MIT}{Massachusetts Institute of Technology}
\newacronym{ECG}{ECG}{Electrocardiogram}
\newacronym{PVC}{PVC}{Ppremature ventricular contraction}
\newacronym{FSVM}{FSVM}{Fuzzy Support Vector Machine}
\newacronym{SICH}{SICH}{Symptomatic Intracranial Haemorrhage}
\newacronym{PCA}{PCA}{Principal Component Analysis}
\newacronym{DNN}{DNN}{Deep Neural Networks}
\newacronym{EEG}{EEG}{Electroencephalography}
\newacronym{ML}{ML}{Machine Leaning }
\newacronym{CART}{CART}{Classification and Decision Tree}
\newacronym{DTW}{DTW}{Dynamic Time Warping}
\newacronym{MRI}{MRI}{Magnetic Resonance Imaging}
\newacronym{ROC}{ROC}{Receiver Operating Characteristic}
\newacronym{PRC}{PRC}{Precision Recall Curve}
\newacronym{TP}{TP}{True Positive}
\newacronym{TN}{TN}{True Negative}
\newacronym{LSTM}{LSTM}{Long-Short Term Memory}

\journal{Medical \& Biological Engineering \& Computing}

\newcommand{\etal}{\textit{et al}.}

\begin{document}

\begin{frontmatter}

\title{Predictive and diagnosis models of stroke from hemodynamic signal monitoring}

\author[1]{Luis García-Terriza}
%\cormark[1]
\author[1]{José L. Risco-Martín}
\author[2]{Gemma Reig Roselló}
\author[1]{José L. Ayala}
\address[1]{Dpt. of Computer Architecture and Automation, Complutense University of Madrid, Spain}
\address[2]{Stroke Care Unit, Hospital Universitario de La Princesa, Spain}

%\begin{graphicalabstract}
%\begin{center}
  %\makebox[\textwidth]{\includegraphics[width=0.8\paperwidth]{graphical_abstract.png}}
%\end{center}
%\end{graphicalabstract}

\begin{abstract}
This work presents a novel and promising approach to the clinical management of acute stroke. Using machine learning techniques, our research has succeeded in developing accurate diagnosis and prediction real-time models from hemodynamic data. These models are able to diagnose stroke subtype with 30 minutes of monitoring, to predict the exitus during the first 3 hours of monitoring, and to predict the stroke recurrence in just 15 minutes of monitoring. Patients with difficult access to a \acrshort{CT} scan, and all patients that arrive at the stroke unit of a specialized hospital will benefit from these positive results.
The results obtained from the real-time developed models are the following: stroke diagnosis around $98\%$ precision ($97.8\%$ Sensitivity, $99.5\%$ Specificity), exitus prediction with $99.8\%$ precision  ($99.8\%$ Sens., $99.9\%$ Spec.) and $98\%$ precision predicting stroke recurrence  ($98\%$ Sens., $99\%$ Spec.).

\makebox[\textwidth]{\includegraphics[width=0.83\paperwidth]{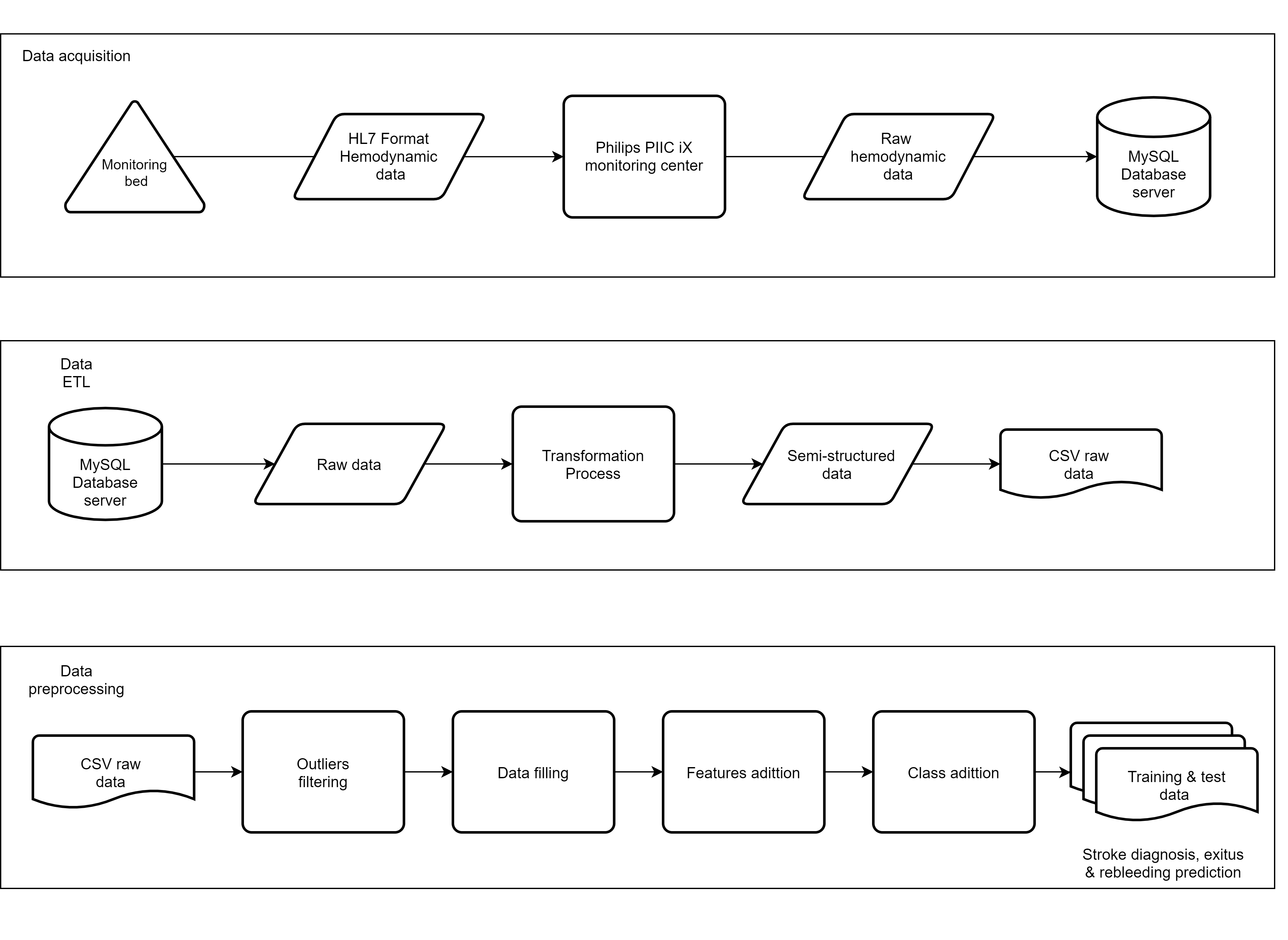}}

Graphical abstract depicting the complete process since a patient is monitored until the data collected is used to generate models

\medskip
\noindent Total words of the manuscript: 5694 words
\newline
The number of words of the abstract: 124
\newline
Number of figures: 5
\newline
Number of tables: 12

\end{abstract}


\end{frontmatter}

%\linenumbers

\section{Introduction}
Stroke is a cerebrovascular disease that has great health and social impact due to its high incidence and prevalence. It is one of the main death causes in the world and produces serious long-term disabilities. %Stroke constitutes the first cause of acquired disability in adults and the second cause of dementia after Alzheimer's disease \cite{pmid24238835}. 
It is a great burden, not only from the health point of view but also personal, family and health/social cost, because of its impact on peoples life's, patients and caregivers \cite{Stevens:17}. 
%In Spain, stroke is a very common cause of morbidity, mortality, and hospitalization. It is the third leading cause of mortality in the general population, and the first in women (3). Stroke causes very high health and social cost, which is estimated to increase in the coming years, due to the aging of our population.

%Health and social cost comes not only directly from the acute phase and hospitalization time, but also from the long term rehabilitation process and the need of patient's permanent institutionalization in the worst cases.

There are different stroke subtypes, what implies different clinical management and therapeutic approaches. Intracerebral hemorrhage needs early admission in the stroke unit for blood pressure monitoring. All stroke patients need early evaluation by a neurologist to have access to prompt reperfusion therapy. Large vessel occlusion stroke needs prompt neurointerventional evaluation. 

Acute stroke management implies a complex healthcare organization. %In Madrid, there is a stroke care plan (Plan de Atención al Ictus en la Comunidad de Madrid), that defines the whole process from pre-hospital care to the final rehabilitation center. 
In the acute phase, pre-hospital emergency units are aware and trained to detect a stroke. %Once a stroke is detected, the stroke code starts. The pre-hospital unit calls the emergency center coordinator which contacts directly to the neurologist on call. 
Depending on the clinical features, patients will be sent to the closest stroke unit hospital or stroke center (if large vessel occlusion is suspected). %In order to detect large vessel occlusion, in a pre-hospital setting, clinical scales are used such NIHSS and Madrid-direct scales. 
Once patients arrive at the closest and correct hospital, a neurologist is waiting in the emergency box. A neurological, cardiological and respiratory examination is performed, and fast transfer to the CT scan is done. %After CT-scan, a precise diagnosis can be provided. In cases of a cerebral hemorrhage, the patient will be transferred to ICU (intensive care unit) or stroke unit, depending on severity. If an ischemic stroke is diagnosed, further evaluation will be needed. After evaluation of the performed acute reperfusion treatment, patients will be mostly transferred to the stroke unit. During the stay in the stroke unit, patients will be semi-intensive monitored (ECG, pulse oximetry, respiratory frequency, heart rate, blood pressure). Once the patient is stabilized, transfer to a general hospitalization bed will be possible.

A stroke unit is an organized in-hospital facility that is fully devoted to care for patients with stroke. %It is staffed by a multidisciplinary team with special knowledge in stroke care. 
The semi-intensive monitoring system during acute stroke phase, and a well trained team, allows to detect and sometimes control all possible neurological complications, which depending on subtype stroke diagnosis are: i) rebleeding in cerebral hemorrhage; ii) recurrence and/or hemorrhagic transformation in ischemic stroke cases.

During the stay of the patients in the stroke units, patients are monitored for a long time, generating a large amount of heterogeneous data that, currently, are not exploited to help on the diagnosis, evolution, or treatment of the patients.  

Applying machine learning methods and algorithms over biomedical records is nowadays providing remarkable results. %Data gathering from patients has improved considerably in the last years, as well as preprocessing and algorithms used to perform predictions. 
Machine learning ensures unbiased decisions that are only made according to the state of the patient, profiled by the gathered data. Machine learning is more than a suitable option for performing predictions as patients are monitored continuously since the hospital admission, and data are stored ordered and structured in a database. %Monitored data are considered heterogeneous, that is to say, every observation of each patient is considered equally valid as a state of the patient, which allows performing training-testing-validation methods with all the data gathered. 

There are more cases where machine learning has been used to aid or improve the diagnosis or detection of diseases, like Parkinson's disease \cite{oh2018deep}, or to predict the results of coronary angiography \cite{groselj1997machine}.

%\cite{oh2018deep} applies deep learning algorithms to aid in the detection of Parkinson's disease from an electroencephalogram (EEG) signal, obtaining $88.25\%$ of accuracy. \cite{groselj1997machine} aimed to predict the results of coronary angiography by using machine learning classic algorithm, Naive Bayesian Classifier, training with all data available (signs, symptoms, ECG signals, MPS). They obtained slightly better results compared to physicians, 88\% versus 84\%; according to \cite{groselj1997machine}, these results could mean that potentially $15.6\%$ less patients would not have to be examined with other tests in order to confirm the diagnosis.

In this research, we develop and apply machine learning algorithms and methods to perform predictions over data from patients who have suffered a stroke. Stroke monitoring data have a great temporal component, what is a difficult factor to deal with in machine learning and deep learning analysis. There is a challenge as well with the balance of classes, as every predictive target is totally off-balance. This may incur in an over-fitting and bad sensitivity/specificity. Therefore, it must be considered and managed from the problem conception.

Being able to predict stroke subtype in a hospital setting, especially when brain CT-scan is possible, may not seem very useful. However, this result is extremely interesting to be applied in other settings like rural areas that are far away from local hospitals, which not always have access to CT image. In these cases, a prompt screening of the patients is mandatory. Additionally, the prediction of acute neurological complications, such as rebleeding, recurrence or hemorrhagic transformation, could allow health professionals to react in time in order to avoid them or at least minimize adverse effects. Once in the hospital, being able to predict death not only gives opportunities to doctors to change the therapeutic approach but, in cases that this is not possible, giving an accurate prognosis risk to patients and/or families is relevant.

The goals of our research are:

\begin{itemize}
    \item efficient preprocessing of real-time data to increase the quality of the acquired data and allow knowledge extraction;
    \item development of machine-learning approaches for modeling of stroke subtype, exitus caused by stroke, and recurrence of the episode;
    \item selection of best model strategy, and validation of results from a computational and clinical perspective.
\end{itemize}

In the following section the state of the art is related, where the current techniques, methods, researches and approaches are presented and explained. Next, the methodology section will present the methods, techniques and  procedures followed in this research. Finally, results and clinical conclusion will end this article showing the results obtained and the clinical and computational discussion.

\section{Approach}

Recent research in the field of artificial intelligence has been incorporated into the clinical field, targeting neurological diseases like stroke. 

Some works have aimed to predict strokes by modeling over demographic and public datasets. \cite{khosla2010integrated} used the \acrfull{CHS} dataset to predict eventual stroke by modeling \acrfull{SVM}.%, obtaining better results with respect to the Cox model technique. 
\cite{letham2013interpretable} developed a generative model called  Bayesian List Machine capable of predicting stroke in atrial fibrillation patients.

Machine learning models and techniques applied to time-series variable data is an area of research in continuous development. \cite{zhang2005ecg} trained a \acrshort{SVM} model by using the \acrshort{ECG} data from the MIT-BIH database that analyzes \acrshort{ECG} and classifies the different sorts of heartbeats, including \acrfull{PVC}. Another research that used the MIT-BIH Arrhythmia Database was \cite{ozcan2010fuzzy}, who developed an \acrshort{FSVM}  (Fuzzy SVM) model to detect Arrhythmia. \cite{garg2019automating} sought to use natural language processing techniques of electronic health records to determine ischemic stroke sub-type treatment.%, concluding that automated machine learning algorithms show agreement with manual acute stroke treatment classification.

Brain image analysis is currently the state-of-art in stroke sub-type diagnosis according to \cite{aviv2007early}. AI in the brain image area has been deeply developed in the last years, obtaining relevant results (\cite{lee2017deep}). There have been researches that apply machine learning algorithms on imaging data to predict or diagnose outcomes related to strokes. \cite{bentley2014prediction} used \acrshort{SVM}  models to predict \acrshort{SICH} outcomes by using CT images of patients who suffered an acute ischemic stroke. \cite{cheon2019use} sought to predict the stroke itself by applying \acrshort{PCA}  plus \acrlong{DNN}  by using medical service data and health behavior data from more than $15000$ patients. %They compared their machine learning method (PCA + DNN) to other traditional machine learning models, obtaining AUC (Area Under Curve) metrics above $80\%$. 

\cite{asadi2014machine} used machine learning algorithms to predict the outcomes of patients after being treated with intra-arterial therapy.  \cite{Zhang2013} applied regression-based algorithms to predict outcomes of the patients three months after suffering the incident by using data within the first 48 hours. \cite{vrtkova2019comparing} aims to predict outcome out of a Rankin Scale after stroke by comparing the performance of different machine learning methods like Regression models, Random Forests and neural networks, obtaining accuracy near to $90\%$.

Our work, on the other hand, applies machine learning models over hemodynamic, real-time data, for the first time in literature, to achieve several novel goals as stroke sub-type diagnosis, prediction of the eventual death of the patient caused by the stroke, and prediction of stroke recurrence. Previous works have never addressed these goals with such clinical relevance, and have not dealt with the complexity of processing the real-time hemodynamic data. Our results not only represent significant advances in the field of applied machine learning techniques in the biomedical field, but also open new therapeutic perspectives in the management of stroke.

\section{Methods}

\subsection{Dataset}

The dataset used in this study was acquired with a Philips PIIC iX monitoring center installed in the stroke care unit of the Hospital Universitario La Princesa. Hemo-dynamic data collection was performed from Spring 2017 until October 2019, and 798 patients were successfully monitored. 

Patients were monitored from the moment they were admitted to the Stroke Care Unit until they were transferred to the observation floor. All patients admitted in the stroke unit were monitored. Inclusion criteria for stroke unit monitoring included all types of stroke in the acute phase (mostly within 24-48 hours after the stroke onset), but longer monitoring is possible if close neurological and semi-intensive monitoring is needed. No age limit is used as exclusion criteria. The only exclusion criteria for monitoring is immediate death or previous severe dependence (modified ranking scale mRS 4-5, GDS - Global Deterioration Scale - 6-7). Some patients with mRS 4 due to musculoskeletal disease but good cognitive state are also admitted. Before stroke care unit admission, patients are evaluated by a neurologist on call in the emergency room (mostly in emergency box). A CT scan is performed. After ruling out hemorrhage, in the cases that an ischemic stroke is suspected, a perfusion CT and an angio-CT (cerebral and supraortic vessels) is performed. In the cases of intracerebral hemorrhage, a post-contrast CT is done, in order to evaluate spot sign and evaluate active bleeding and rebleeding risk. In cases that a stroke mimic is suspected, the patient will be transferred to the emergency care unit and more tests (\acrshort{EEG}, lumbar puncture,...) will be done to achieve a final diagnosis. The information recorded from the patients is based on non-invasive measures of hemodynamic variables, as described in Table \ref{tab:variables}. Timestamps for the acquired data are also considered in the analysis, as well as demographic information like age and gender. 

The clinical variables that were registered are the type of stroke (ischemic vs hemorrhage), type of hemorrhage (intraparenchymal vs non), death and time to death since admission, time to any acute neurological complication that includes early recurrence stroke, hemorrhagic transformation in acute ischemic stroke, and rebleeding in acute intraparenchymal hemorrhage. Exclusion criteria after monitoring analysis were patients with non-stroke final diagnosis (i.e. seizures, CNS infection, migraine, ...) or scheduled hospital admission to perform carotid angioplasty or aneurysm embolization. 

The monitoring center stores all the information in a relational database, where each variable measured compounds a row in the relational database. Each row is structured as follows: variable ID, variable value, variable unit, timestamp, and patient ID. Therefore, several rows must be preprocessed together to obtain a complete observation of the patient. Observations do not include demographic variables yet because, as previously mentioned, training instances are compounded by several observations. This observation is acquired and stored in the database every $30$ seconds.

One observation may provide a significant amount of information from the patient as it represents the state of the patient for 30 seconds; but in a time series prediction or diagnosis where the patient is monitored for hours, it is not enough for machine learning algorithms to provide reliable predictions or diagnosis. Henceforth, we have put together several consecutive observations shaping a training or testing instance. The number of observations per instance was a relevant part of the study, where the minimum number of observations compounding an instance were five observations (what represents $2.5$ minutes of monitoring time) and the maximum number was one hundred and twenty observations (what represents $1$ hour of monitoring time). In this study, the number of observations for all the results is five, as it was the time interval that performed best. Training instances do include as well demographic variables (age, gender) and the timestamp of the first observation measured.

From the original dataset that includes 798 patients, several datasets have been created as required by the experimental work. This happens because diagnostics were not available for all of the patients by the time these experiments were performed. In the case of stroke subtype diagnosis, the dataset includes 468 ischemic patients and 80 hemorrhagic patients. Regarding exitus statistics, 43 out of 504 passed away due to stroke during their hospitalization, and 34 out of 500 suffered a stroke recurrence episode during their hospitalization.

\begin{table*}
\caption{Definition of variables used in this study}
\centering
\begin{tabular}{ccccc}
\hline
Variable & Abbreviation & Type & Role & Description\\
\hline
Type of Stroke & TS & Categorical & Ischemic/Hemorrhagic & Target \\
Risk Prediction & RP & Categorical & Exitus/Non Exitus & Target \\
Rhythm Estimation & RE & Numerical & Input & A Rhythm Indicator \\
VE & VE & Numerical & Input & Ventricular extra systole\\
CF & CF & Numerical & Input & Cardiac Frequency \\
Breathing Frequency & BF & Numerical & Input & Respiratory Rate \\
Perf & - & Numerical & Input & Pulmonar Perfusion \\
SpO2 & - & Numerical & Input & Oxygen Saturation \\
ST-II & - & Numerical & Input & Syst. Time Interval Index \\
\hline
\end{tabular}
\label{tab:variables}
\end{table*}

\subsection{Preprocessing}

%Preprocessing is crucial in this kind of experimental setups as variables are acquired by sensors in a clinical environment where wrong measurements and incoherence are likely due to bad connections or punctual malfunctioning. 

\begin{figure*}
    \centering
    
    \includegraphics[width=\textwidth]{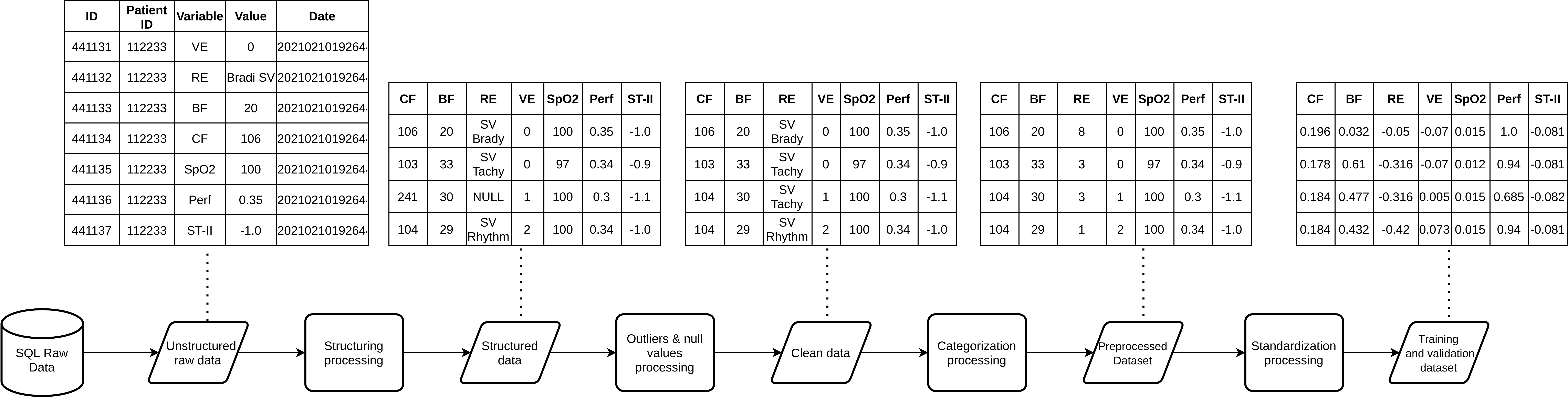}
    \caption{Data preprocessing: data cleaning and imputation phases}
    \label{fig:preprocessing}
\end{figure*}

The first preprocessing phase consisted of removing outlier values, being replaced by the last correct value or the mean of the previous recordings, depending on the variable type. In the case of numerical data (SPO2, EV, etc.) the value was replaced by the mean of the previous recordings of that patient when the value was greater or equal to four times the standard deviation. On the other hand, categorical values (Rhythm Estimation) were replaced by the last value when the value was not measured correctly. The same procedure was followed for missing values.

This preprocessing phase is graphically depicted in Figure \ref{fig:preprocessing}.

Finally, the last preprocessing phase consisted on standardizing data. As each feature has different ranges and different measurement units, data must be standardized in order to prevent unbalance in the algorithm's internal weights. Therefore, Z-Score standardization was performed for every feature, where each value is subtracted the mean value ($\mu$) and then that result is divided by the standard deviation ($\sigma$):

\begin{equation}
\mathrm{Z} = \frac{X-\mu}{\sigma}
\end{equation}

Signal preprocessing was not performed since that task was efficiently done by the monitoring center.

\subsection{Algorithms}

One of the main goals of this study was to discover which learning algorithms achieve the best performance in terms of prediction capability. Initially, we evaluated at least one algorithm for each sort of \acrshort{ML} model but, after several tests, some of them were discarded due to their bad performance. The criteria to discard algorithms was as follows: each algorithm was compared against the best performing algorithm and, if the F-Measure difference was more than $10\%$, the algorithm was discarded. Discarded algorithms were Logistic Regression, SVM, Naive Bayes, Multi-Layer Perceptron Neural Network, Long-Short Term Memory Deep Neural Network, and Convolutional 1-D Deep Neural Network. 

Thus, the algorithms chosen and shown in this research are the following:
\begin{itemize}
    \item \textbf{Decision Tree}: a non-parametric model that create a model targeting the value of each variable by simple decision rules inferred from data. The optimal parameters found were 17 as maximum depth, \textit{gini} as split function, and a minimum of 2 samples to split a node. 
    \item \textbf{Random Forests}: a ensemble algorithm typically used for classification and regression. It fits several classifiers (usually decision trees), also known as estimator, and the output of the algorithm is obtained by taking the most repeated output of the classifiers. It performs especially well because it avoids over-fitting due to the way dataset is distributed between the indicators. Optimal parameters were $1000$ estimators, maximum depth of $23$ and \textit{gini} as entropy algorithm. Rest of parameters were set to their default values. 
    \item \textbf{Gradient Boosting}: a ensemble algorithm that builds the model in stages, also known as estimators. In each stage, it builds $N$ regression trees and finally, it generalizes all of them by allowing optimization of an arbitrary loss function (e.g., Deviance Loss Function). The optimal parameters were 100 estimators, maximum depth of 5, $1.0$ as learning rate and \textit{Deviance Loss} as loss metric. The rest of the parameters were set as default.
    \item \textbf{AdaBoost}: a meta-estimator that builds the model by fitting a classifier (typically a \acrshort{CART} Decision Tree Classifier) and comparing to another fitted classifier, where weights of incorrectly classified examples are adjusted such that subsequent classifiers are focused on wing cases. Optimal parameters were $400$ estimators, $1.0$ as learning rate, \textit{Decision Trees Classifier} as base estimator and \textit{SAMME.R} as base algorithm.
    \item \textbf{K-Nearest Neighbors}: a non-parametric and distance-based algorithm where input consists the base knowledge, i.e, training examples. When K-NN is used for classifying, each output is chosen by the closest \textbf{K} examples of the base knowledge, calculating the distance with the chosen distance metric algorithm. The optimal parameters were $K=17$, \textit{Ball Tree} as compute algorithm and \textit{Minkowski distance, P=2} as distance metric.
    \item \textbf{Dynamic Time Warping}: an algorithm that measures the distance between two time series signals. This algorithm used with a simple K-Nearest-Neighbors with $K=1$ and adapted to multivariate (i.e. applied to measure the distance to several features) it may be used on classification problems. There are two types of multivariate \acrshort{DTW} algorithms: dependent and independent (\cite{shokoohi2017generalizing}). Since dependent \acrshort{DTW} has provided slightly better results than the independent algorithm it is the method shown in this research.

\end{itemize}

\subsection{Experimental evaluation}

In our experimental work, we have performed three sets of experiments related to stroke events. The first consists on diagnosing the type of stroke suffered by the patient (ischemic vs hemorrhagic) within a short temporary window (less than 60 minutes). As the treatment of both types of stroke is totally opposite, an early guess of the event is desired for proper management. It is known that stroke diagnosis gold standard procedure is a brain image scan (\acrshort{CT} or \acrshort{MRI}). But in country areas, having access to a CT is difficult, even more to MRI studies. Most rural medical centers will not have advanced imaging techniques, and alternatives are desired to avoid delays in treatment options. 

The second experiment is to predict a patient's death after stroke, according to incremental time windows from the event. The initial results of these two goals were already published in \cite{garcia2019comparison}. From that preliminary work, we have improved the data pre-processing, added demographic variables, and extended the refinement of machine learning algorithms. These improvements, and a larger dataset, have allowed us to obtain better performance in the algorithms. Death prediction allows clinicians to adopt a different therapeutic approach in order to prevent, and if not possible minimize, any possible future and immediate adverse event due to any clinical complication (neurological, respiratory, cardiological, ...). Additionally, being able to give this information to patients and family is also relevant in clinical practice as relatives require information about risks and prognosis.

Finally, the third experiment consists on predicting any neurological complication of a stroke patient during acute setting monitoring, which includes: early recurrence in ischemic stroke, hemorrhagic transformation in ischemic stroke and rebleeding in intraparenchymal hemorrhage. This is a critical situation for the patient since it drastically decreases the chances of survival. So far, clinicians have not been able to understand the mechanisms that drive a patient into recurrence or rebleeding, and our work is a novel and promising advance in that direction. The development of a model that can predict neurological complications of patients allows the medical staff to take preventive measures to avoid the event or to reduce its severity. 

\begin{figure}
    \includegraphics[width=0.45\textwidth]{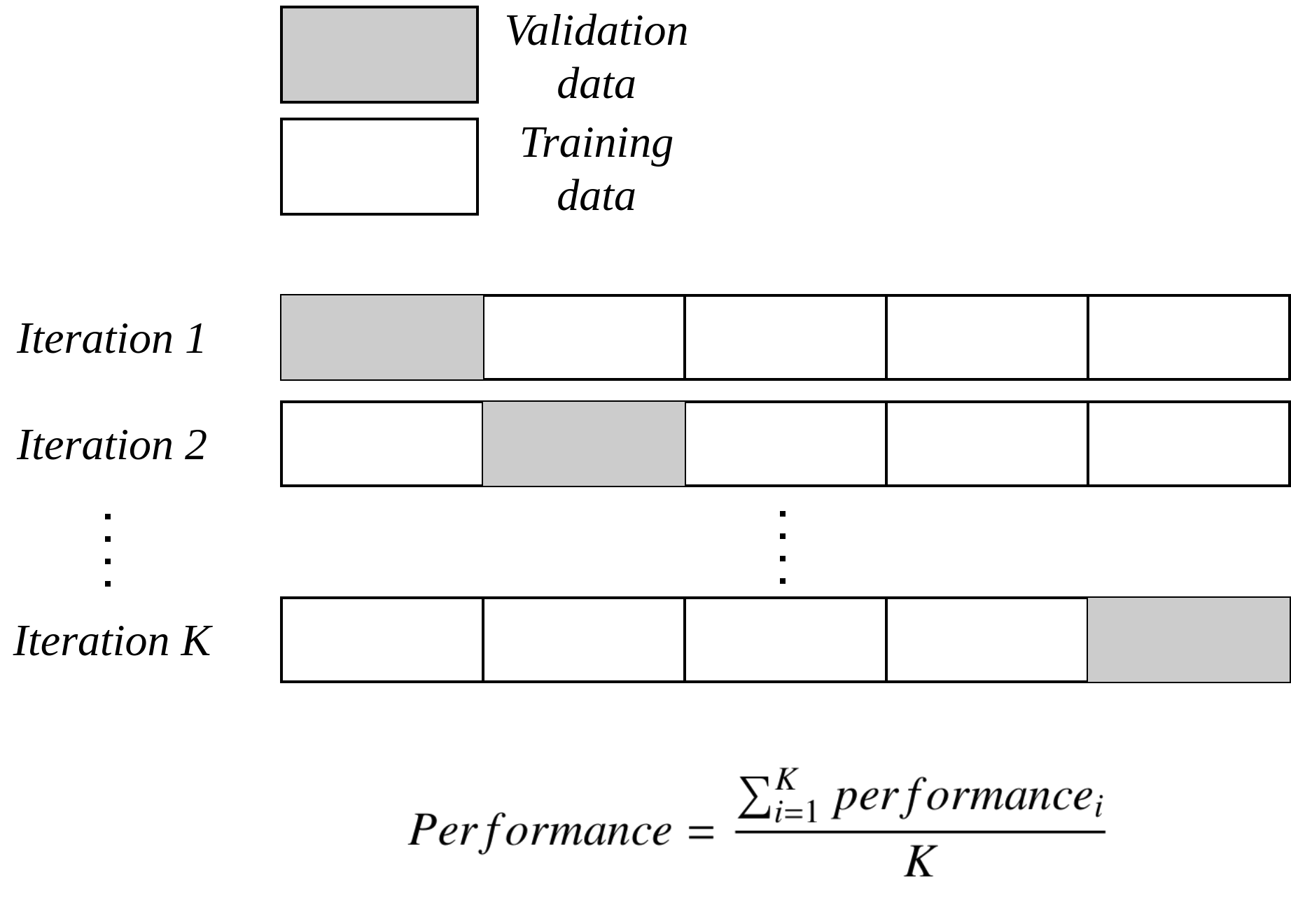}
    \caption{K-Fold cross validation method}
    \label{fig:kfold}
\end{figure}

Our target application demands unbiased results and models objectively evaluated. As depicted in Figure \ref{fig:kfold}, we have performed a K-fold cross-validation methodology with K=5 to avoid over-fitting and obtain results independent from the training and test datasets. The code and libraries used in this study are open source and widely-used: Python with Sci-kit Learn, and Keras. Some of the algorithms used were not available in public libraries repositories, therefore, they were coded for these experiments. The evaluation of the models was measured by the following metrics: accuracy, sensitivity, specificity, F-Measure (F1-Score) and both the areas under the \acrfull{ROC} curve and the \acrfull{PRC}. These metrics are calculated as follows: 

\begin{equation}
\mathrm{Accuracy} = \frac{\mathrm{TP}+\mathrm{TN}}{\mathrm{TP}+\mathrm{TN}+\mathrm{FP}+\mathrm{FN}}
\end{equation}

\begin{equation}
\mathrm{Sensitivity} = \mathrm{Recall} = \frac{\mathrm{TP}}{\mathrm{FN}+\mathrm{TP}}
\end{equation}

\begin{equation}
\mathrm{Specificity} = \frac{\mathrm{TN}}{\mathrm{FP}+\mathrm{TN}}
\end{equation}

\begin{equation}
\mathrm{F-Measure} = \frac{2 \cdot \mathrm{Precision} \cdot \mathrm{Recall}}{\mathrm{Precision}+\mathrm{Recall}}
\end{equation}

\noindent where $\mathrm{Precision} = \mathrm{TP}/(\mathrm{TP+FP})$, $\mathrm{TP}$ is the number of true positives, $\mathrm{TN}$ is the number of true negatives, $\mathrm{FP}$ is the number of false positives, and $\mathrm{FN}$ is the number of false negatives

\section{Results}

As aforementioned, we have performed three different experiments: Stroke type diagnosis, Exitus prediction, and Stroke recurrence prediction. These three experiments were trained, validated and tested with the same initial dataset, being slightly smaller or greater regarding the number of patients as explained before. The dataset features are the same except in the case of Exitus and recurrence predictions where the dataset includes as well the stroke type diagnosed.

Several models were trained for a large distribution of time windows. Before the analysis of the obtained results, we will present how the models (and the time windows) were selected.

\subsection{Model selection}

In order to compare and rank the set of algorithms used, we use the hypervolume indicator $I_H^-$ in the metrics space. This indicator calculates the volume (in the metrics space) covered by members of a non-dominated set of solutions $W$ (\cite{Zitzler1999}). Note that a solution in our case is defined as the set of metrics associated with each time window. For instance, in the case of the stroke sub-type classification, each algorithm has 94 solutions, i.e., 94 time windows with their four metrics. With this in mind, let $v_i$ be the volume enclosed by solution $w_i \in W$ . Then, a union of all hypercubes is found and its hypervolume ($I_H$) is calculated.

\begin{equation}
I_H(W)=\bigcup_1^{\left| W \right|} v_i
\label{eq:HyperVolume}
\end{equation}

The hypervolume of a set is measured relative to a reference point, usually the anti-optimal point or ``worst possible'' point in space. We do not address here the problem of choosing a reference point; if the anti-optimal point is not known or does not exist, one suggestion is to take in each objective the worst value from any of the fronts being compared. If a set $W_i$ has a greater hypervolume than a set $W_j$, then $W_i$ is taken to be a better set of solutions than $W_j$. In this work, we consider the hypervolume difference to a reference set $R$, defined as

\begin{equation}
I_H^-(W)=I_H(R)-I_H(W)
\label{eq:NegativeHyperVolume}
\end{equation}

where smaller values correspond to higher quality. If the reference set is not given, we take $I_H(R)=0$.

\begin{table}[ht]
\centering
\caption{Hypervolumes computed in the case of the stroke sub-type classification system.}
\begin{tabular}{lc}
\hline
Algorithm & HyperVolume \\
Logistic Regression & -0.24 \\
Support Vector Machines(SVC) & -0.17 \\
K-Nearest Neighbour & -1.23 \\
MLP & -0.98 \\
Decision Tree & -1.22 \\
AdaBoost & -1.12 \\
\acrshort{LSTM} & -0.89 \\
Random Forests & -1.38 \\
Naive Bayes & -0.25 \\
Gradient Boosting & -1.46 \\
Dependent Dynamic Time Warping 1-NN & -1.14 \\
\hline
\end{tabular}
\label{tab:HyperVolumes1}
\end{table}

Table \ref{tab:HyperVolumes1} illustrates the set of hypervolumes $I_H^-(W)$ calculated for the stroke sub-type prediction system (with $I_H(R)=0$). As Table \ref{tab:HyperVolumes1}  shows, the best indicator is reached by the Gradient Boosting algorithm, with $I_H^-(W) = -1.46$ followed by Random Forest ($I_H^-(W) = -1.38$) and the K-Nearest Neighbour ($I_H^-(W) = -1.23$). Clearly, according to this indicator the Gradient Boosting algorithm is the best algorithm for stroke sub-type classification.

As stated above, all the hypervolumes are computed over the non-dominated set of solutions, or in other words, the non-dominated set of time windows. A solution $w_i \in W$ is said to dominate another solution $w_j \in W$ (denoted as $w_i \prec w_j$) if the following two conditions are satisfied. 

\begin{eqnarray}
 & \forall i \in \left\{ 1, 2, \ldots , m \right\}, f_{i} \left(  w_i \right) \geq f_{i}  \left( w_j \right) & \nonumber \\ 
 & \exists i \in \left\{ 1, 2, \ldots , m \right\}, f_{i} \left(  w_i \right) > f_{i}  \left( w_j \right) &
\end{eqnarray}

A solution $w$  is non-dominated if another $w'$ such that $w' \prec w$ does not exist. Hence, the set of non-dominated solutions of each algorithm will give us the best time windows to be used in the prediction or the classification system. This is usually known as the Pareto front or more precisely, an approximation to the Pareto front.

\begin{table}[ht]
\caption{Non-dominated solution set of Gradient Boosting and their respective metric values.}
\centering
\resizebox{\columnwidth}{!}{\begin{tabular}{lcccc}
\hline
Time Window & F1-Score & Specificity & Sensitivity & Accuracy \\
5-(0,30) & 0.96 & 0.98 & 0.96 & 0.97\\
5-(0,65) & 0.99 & 1.0 & 0.98 & 0.99\\
5-(0,110) & 0.99 & 1.0 & 0.99 & 0.99\\
5-(0,115) & 0.99 & 1.0 & 1.0 & 1.0\\
5-(0,120) & 0.99 & 1.0 & 0.98 & 0.99\\
\hline
\end{tabular}}
\label{tab:NonDominated1}
\end{table}

Table \ref{tab:NonDominated1} illustrates the non-dominated solution set for stroke sub-type classification with the Gradient Boosting algorithm. As can be seen, the best time window candidate to be used for the classification system remains within the one or two first hours after the patient was admitted in the Stroke Care Unit with only 5 observations per instance. However, a very good result is also obtained for the observations during the first 30 minutes, achieving a much more useful clinical result.

Table \ref{tab:HyperVolumes2} illustrates the set of hypervolumes calculated for the exitus prediction system, where Random Forest and Gradient Boosting again achieve similar results ($I_H^-(W) = -1.46$). 

\begin{table}
\centering
\caption{Hypervolumes computed in the case of the exitus prediction system.}
\begin{tabular}{lc}
\hline
Algorithm & HyperVolume \\
Logistic Regression & -0.68 \\
Support Vector Machines(SVC) & -1.35 \\
K-Nearest Neighbour & -1.37 \\
MLP & -1.44 \\
Decision Tree & -1.45 \\
AdaBoost & -1.42 \\
LSTM & -1.16 \\
Random Forests & -1.46 \\
Naive Bayes & -0.67 \\
Gradient Boosting & -1.46 \\
Dependent Dynamic Time Warping 1-NN & -1.16 \\
\hline
\end{tabular}
\label{tab:HyperVolumes2}
\end{table}

Tables \ref{tab:NonDominated21} and \ref{tab:NonDominated22} illustrate the non-dominated solution set for exitus prediction with the Random Forest and Gradient Boosting algorithms, respectively. As can be seen, the results for the exitus prediction are highly positive from a clinical perspective during the first 3 hours and 5 observations per sample (in the case of the Random Forest), although they increase slightly in the next hours. In the case of the Gradient Boost, results are pretty similar and the same window could be used. 

\begin{table}[ht]
\centering
\caption{Non-dominated solution set of Random Forests and their respective metric values.}
\resizebox{\columnwidth}{!}{\begin{tabular}{lcccc}
\hline
Time Window & F1-Score & Specificity & Sensitivity & Accuracy \\
5-(0,180) & 0.99 & 1.0 & 0.98 & 0.99\\
5-(0,360) & 0.99 & 1.0 & 0.99 & 1.0\\
5-(0,540) & 1.0 & 1.0 & 0.99 & 1.0\\
5-(0,2880) & 1.0 & 1.0 & 1.0 & 1.0\\
5-(0,3600) & 1.0 & 1.0 & 1.0 & 1.0\\
\hline
\end{tabular}}
\label{tab:NonDominated21}
\end{table}

\begin{table}[ht]
\centering
\caption{Non-dominated solution set of Gradient Boost and their respective metric values.}
\resizebox{\columnwidth}{!}{\begin{tabular}{lcccc}
\hline
Time Window & F1-Score & Specificity & Sensitivity & Accuracy \\
5-(0,180) & 0.99 & 0.99 & 0.99 & 0.99\\
5-(0,360) & 0.99 & 1.0 & 1.0 & 1.0\\
5-(0,3600) & 1.0 & 1.0 & 1.0 & 1.0\\
5-(0,4320) & 1.0 & 1.0 & 1.0 & 1.0\\
\hline
\end{tabular}}
\label{tab:NonDominated22}
\end{table}

Similar approach has been followed for the stroke recurrence in Tables \ref{tab:HyperVolumes3} and \ref{tab:NonDominated3}, where DTW 1-NN algorithm has been selected with a time window of 30 minutes and 5 observations per instance.

\begin{table}[ht]
\centering
\caption{Stroke recurrence}
\begin{tabular}{lc}
\hline
Algorithm & HyperVolume \\
Logistic Regression & -0.61 \\
Support Vector Machines(SVC) & -0.01 \\
K-Nearest Neighbour & -1.38 \\
MLP & -1.24 \\
Decision Tree & -1.36 \\
AdaBoost & -1.4 \\
LSTM & -1.23 \\
Random Forests & -1.44 \\
Naive Bayes & -0.46 \\
Gradient Boosting & -1.44 \\
Dependent Dynamic Time Warping 1-NN & -1.46 \\
\hline
\end{tabular}
\label{tab:HyperVolumes3}
\end{table}

\begin{table}[ht]
\centering
\caption{Non-dominated solution set of Dependent Dynamic Time Warping 1-NN and their respective metric values.}
\resizebox{\columnwidth}{!}{\begin{tabular}{lcccc}
\hline
Time Window & F1-Score & Specificity & Sensitivity & Accuracy \\
5-(0,10) & 0.92 & 0.963 & 0.8846 & 0.9245 \\
5-(0,30) & 0.9505 & 0.9896 & 0.923 & 0.9662 \\
5-(0,50) & 0.9090 & 0.939 & 0.946 & 0.941 \\
\hline
\end{tabular}}
\label{tab:NonDominated3}
\end{table}

\subsection{Performance Results for Stroke Diagnosis Models}

\begin{table*}
\centering
\caption{Stroke diagnosis model: performance metrics}
\begin{tabular}{cccccc}
\hline 
Perf. metric & DTW & Nearest Neighbour & Gradient Boost & Random Forests & Decision Tree \\
Sensitivity & 0.8736 & 0.9134 & \textbf{0.9783} & 0.9567 & 0.8881 \\
Specificity & 0.9509 & 0.9701 & \textbf{0.9957} & 0.9893 & 0.9594 \\
F-Measure & 0.8930 & 0.9301 & \textbf{0.9855} & 0.9689 & 0.9077 \\
Accuracy & 0.9221 & 0.9490 & \textbf{0.9893} & 0.9772 & 0.9329 \\
ROC Area & 0.9123 & 0.9417 & \textbf{0.9870} & 0.9730 & 0.9237 \\
PRC Area & 0.9169 & 0.9794 & \textbf{0.9994} & 0.9975 & 0.9290 \\
Avg & 0.9115 & 0.9473 & \textbf{0.9892} & 0.9771 & 0.9235 \\
\hline
\end{tabular}
\label{tab:diagnosis}
\end{table*}

\begin{figure}
\centering
\includegraphics[width=0.5\textwidth]{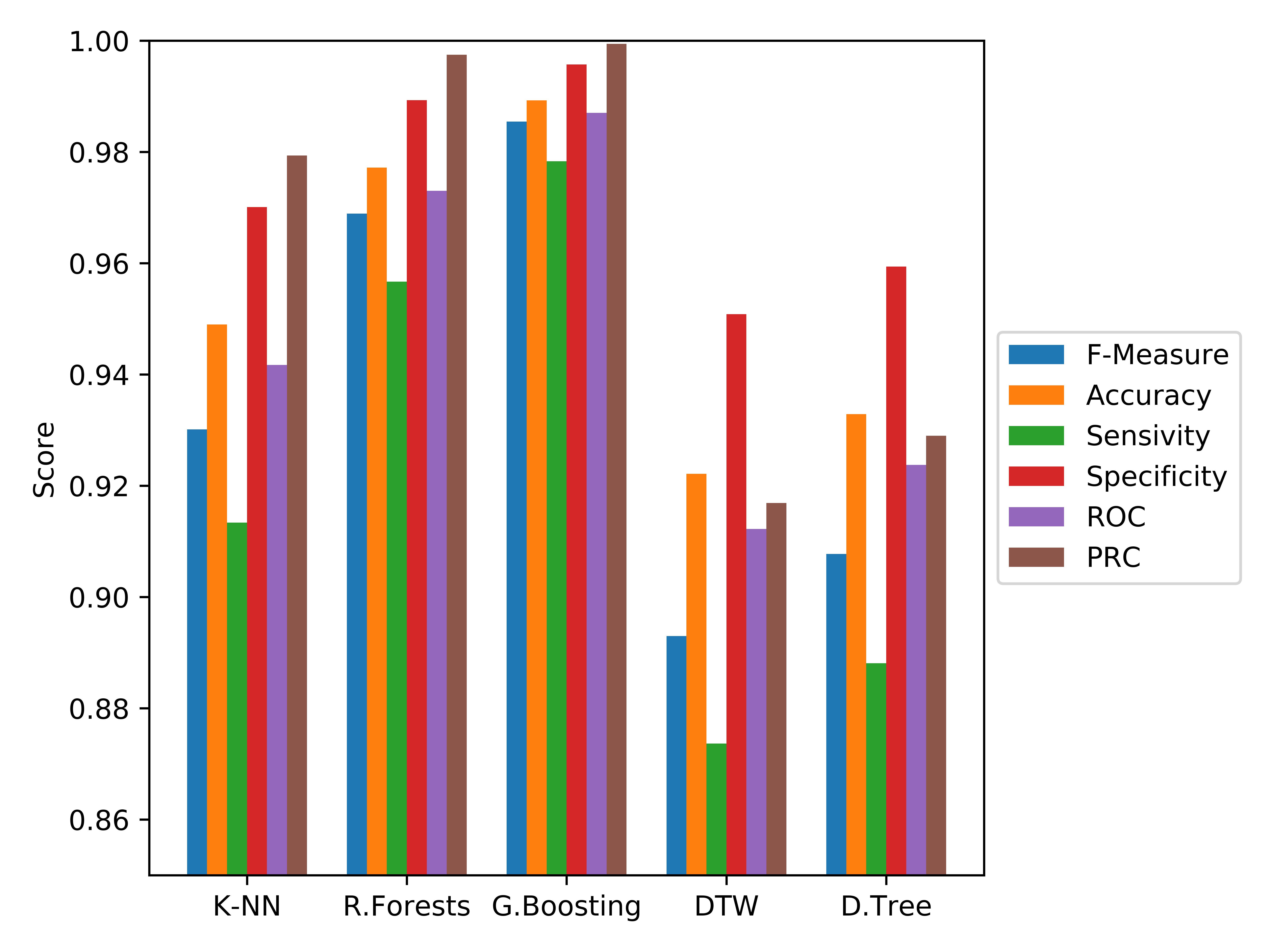}
\caption{Diagnosis detection: all the performance metrics of each model together.}
\label{fig:diagnosis}
\end{figure}

In this set of experiments, the aim is not to classify for making a prediction but to achieve a diagnosis or detection of the stroke type. As aforementioned, the stroke can be either ischemic or hemorrhagic and detecting the type of stroke is crucial for the proper treatment. 

The results presented in this section were obtained after training and testing with data from 548 patients, where 468 patients were ischemic and the remaining 80 patients were hemorrhagic. The tests were performed through several temporal windows and the results shown here correspond to data within the first 30 minutes of monitoring, which is the minimum time required to obtain these results. Extending the monitoring time does not improve results, and reducing the monitoring time decreases scores markedly. The first tests showed that the unbalance of classes was causing an under-fit of the smallest class hence we decided to balance classes to a proportion where the smallest class formed the 35\% of the dataset.

Table \ref{tab:diagnosis} shows the detailed results of the performance metrics. The last row of the table shows the average of the rest of the metrics. The F-Measure results were $0.9855$ for Gradient Boosting,  $0.9689$ for Random Forests, $0.9077$ for Decision Tree, $0.8930$ for Dynamic Time Warping and $0.9301$ for Nearest Neighbors. As can be seen, Gradient Boosting algorithm obtained the best results on each metric. The F-Measure score $0.9855$ is notably higher than results achieved in previous work \cite{garcia2019comparison} where Random Forests performed better. On the other hand, the algorithm that provided the lowest results was \acrlong{DTW}. The reason for this behavior can be probably explained by the observation time (5 observations per instance), that is too short for the proper performance of DTW. 

These results can be seen graphically in Fig. \ref{fig:recurrence}.

Being able to detect hemorrhage vs cerebral infarct within only 30 minutes monitoring allows to apply the correct treatment and/or transfer to the right medical center (stroke unit vs stroke center) which implies the possibility to receive early and accurate treatment depending on each medical case. Therefore, most patients from rural areas, in extra-hospital emergency care or medical centers without CT scan availability, could benefit from this kind of analysis.

\subsection{Performance Results for Exitus Models}

\begin{table*}
\caption{Exitus prediction: performance metrics}
\centering
\begin{tabular}{cccccc}
\hline 
Perf. metric & DTW & Nearest Neighbour & GradientBoost & AdaBoost & Random Forests \\ 
sensitivity & 0.8757 & 0.9416 & \textbf{0.9981} & 0.9718 & 0.9812 \\ 
Specificity & 0.9407 & 0.9844 & 0.9990 & 0.9825 & \textbf{1.0000} \\ 
F-Measure & 0.8798 & 0.9551 & \textbf{0.9981} & 0.9690 & 0.9905 \\ 
Accuracy & 0.9185 & 0.9699 & \textbf{0.9987} & 0.9788 & 0.9936 \\ 
ROC Area & 0.9082 & 0.9630 & \textbf{0.9986} & 0.9771 & 0.9906 \\ 
PRC Area & 0.9010 & 0.9836 & \textbf{1.0000} & 0.9940 & 0.9999 \\ 
Avg & 0.9040 & 0.9663 & \textbf{0.9988} & 0.9789 & 0.9926 \\ 
\hline
\end{tabular}
\label{tab:exitus}
\end{table*}

The exitus prediction aims to forecast the eventual death of the patient. As previously mentioned, being able to predict the eventual death of the patient gives the medical team the possibility of applying in advance a special treatment and, in some cases, avoid the death of the patient or to increase his survival probability. The methodology in this experiment is similar to the previous experiment in terms of preprocessing and most of the features used, but in this case, the aim is prediction instead of diagnosis.  The dataset employed is composed of monitored data from 504 patients where 43 passed away by stroke causes. Like the previous section, there is a notable unbalance between the classes hence a balance was performed by discarding training and testing instances from the greater class until a balance of 35\%-65\% was achieved.

As exposed in the model selection subsection, the results presented in this study were achieved with a temporal window of the first three hours of monitoring and size of five observations per training instance. 

Being able to predict a patient's death in 3 hours offers an opportunity for health professionals to act. Applying different therapeutic tools to save the patient's life or avoiding acute stroke complications that could lead to death. In cases that it is impossible to avoid death, being able to give certain and accurate prognosis information is very important for patients and families and allows to plan the therapeutic effort.

Table \ref{tab:exitus} presents the results of the performance metrics. The algorithm that performed best was Gradient Boost with $0.9988$ score average, whereas Random Forests achieved $0.9926$, AdaBoost $0.9789$, Nearest Neighbors $0.9663$ and Dynamic Time Warping $0.904$. Gradient Boost and Random Forests predict exitus almost perfect, with very high results in both sensitivity and specificity. 

This experiment was performed beforehand as published in our previous work \cite{garcia2019comparison} with fewer patients and fewer features compounding the model. The results shown in this study are notably better than former results.

\begin{figure}
\caption{Exitus prediction: all the performance metrics of each model together.}
\centering
\includegraphics[width=0.5\textwidth]{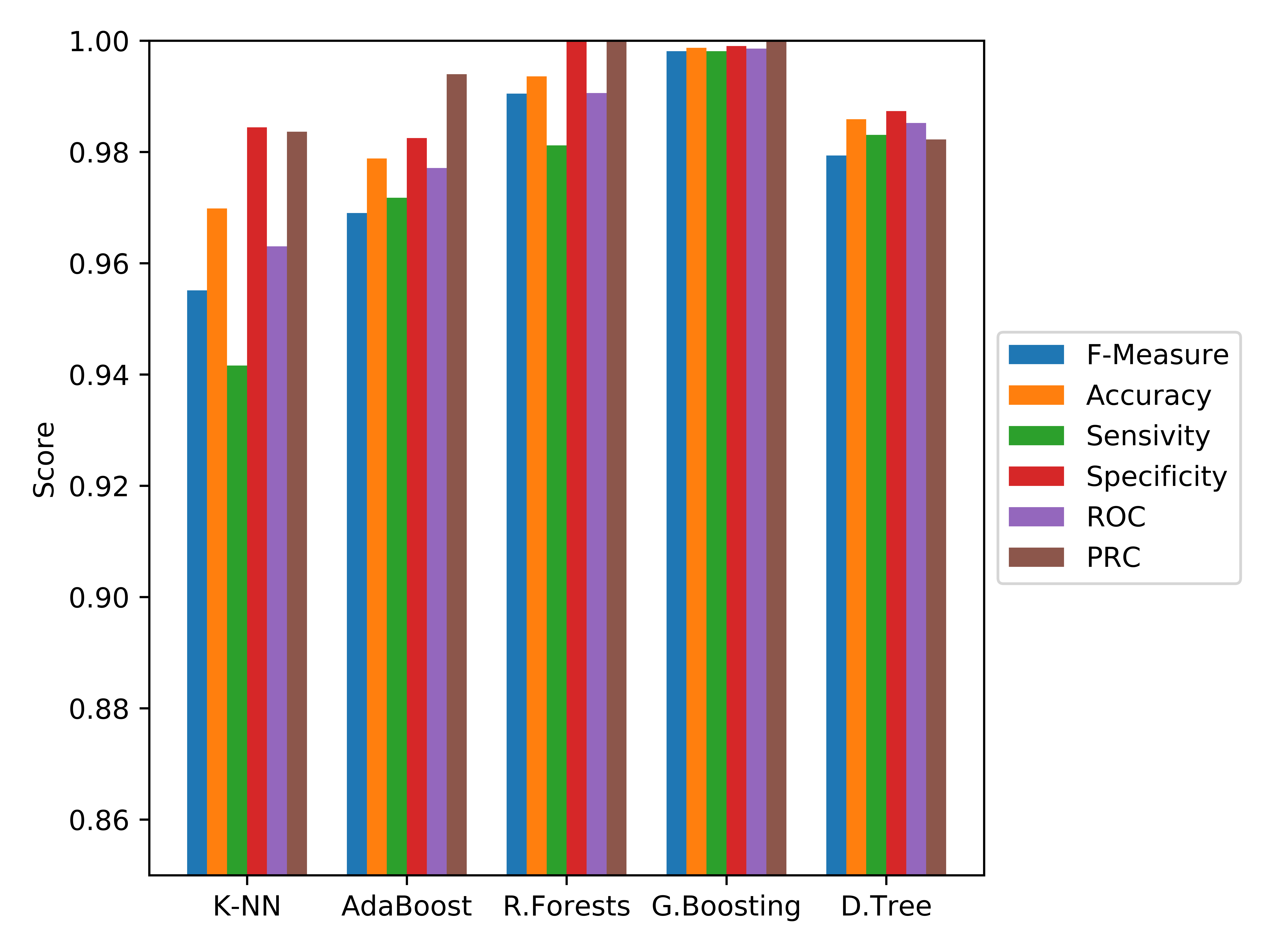}
\label{fig:exitus}
\end{figure}

Fig. \ref{fig:exitus} shows graphically all these results together.

\subsection{Performance Results for Stroke recurrence Models}

\begin{table*}
\centering
\caption{Stroke recurrence prediction: performance metrics}
\begin{tabular}{cccccc}
\hline 
Perf. metric & DTW & Nearest Neighbour & GradientBoost & AdaBoost & Random Forests \\
Sensitivity & 0.9231 & 0.9038 & 0.9231 & \textbf{0.9808} & \textbf{0.9808} \\
Specificity & \textbf{0.9896} & 0.9792 & 0.9583 & 0.9375 & \textbf{0.9896} \\
F-Measure & 0.9505 & 0.9307 & 0.9231 & 0.9358 & \textbf{0.9808} \\
Accuracy & 0.9662 & 0.9527 & 0.9459 & 0.9527 & \textbf{0.9865} \\
ROC Area & 0.9563 & 0.9415 & 0.9407 & 0.9591 & \textbf{0.9852} \\
PRC Area & 0.9648 & 0.9841 & 0.9892 & 0.9919 & \textbf{0.9993} \\
Avg & 0.9584 & 0.9487 & 0.9467 & 0.9596 & \textbf{0.9870} \\
\hline
\end{tabular}
\label{tab:recurrence}
\end{table*}

\begin{figure*}
\centering
\includegraphics[width=0.5\textwidth]{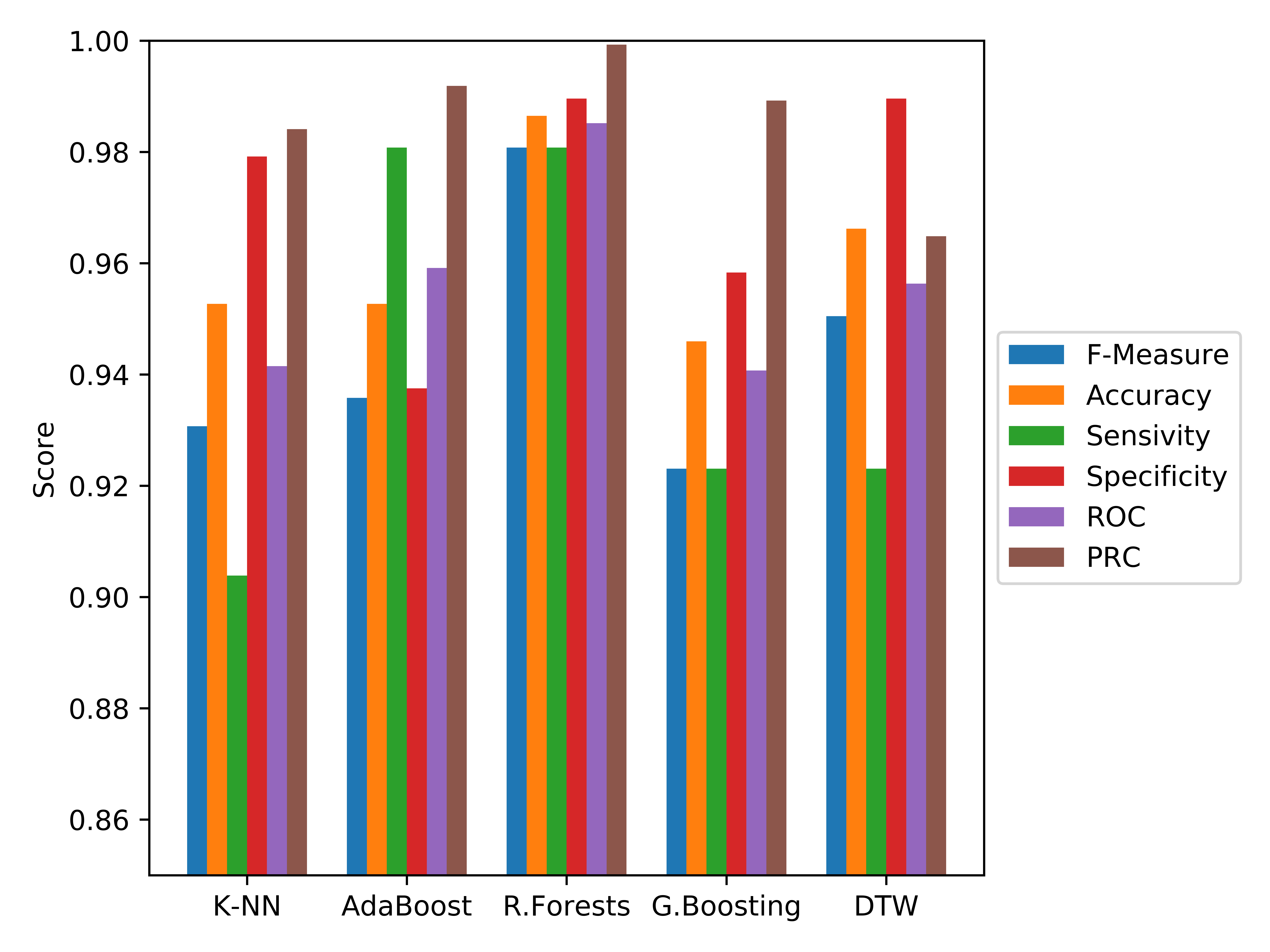}
\caption{Stroke recurrence prediction: all the performance metrics of each model together.}
\label{fig:recurrence}
\end{figure*}

From a dataset composed of 500 patients from which 34 suffered a recurrence during their hospitalization. The remaining 466 patients did not have any kind of recurrence of affection. It can be easily spotted that the weights of classes are not ideal, where only 6.8\% had a recurrence. During our tests, performance decreased substantially when classes were extremely unbalanced. The best way of achieving a model able to predict both classes with high hit rates was to balance classes. Due to this, classes were balanced to a proportion where 35 percent of the observations were from recurrence patients and the remaining 65 percent were from patients who did not have a recurrence during hospitalization. In order to do this task properly,  observations of patients belonging to the majority class were discarded.

Table \ref{tab:recurrence} shows the detailed metric scores for each algorithm tested, where Random Forests is the algorithm that obtains the best results with a $0.98$ percent, followed with $0.9504$ by DTW, $0.9357$ for AdaBoost, $0.9307$ for K-Nearest Neighbors and $0.923$ for Gradient Boosting. Random Forests was the algorithm with best average results and, on the opposite side, DTW obtained the worst results.

The results shown correspond to observations from the first 15 minutes of monitoring of the patients, which is enough to obtain these results. Creating the model with a longer time window provides similar results, hence this is the minimum time window necessary for predicting recurrences on patients.  Predicting any kind of acute stroke neurological complications (i.e.: rebleeding in cerebral hemorrhage, stroke recurrence or hemorrhagic transformation in ischemic stroke) in such a short period of time gives multiple therapeutic opportunities to avoid or at least minimize adverse events. 

Figure \ref{fig:recurrence} depicts the results achieved in this experiment together.

\begin{table*}
\centering
\caption{Comparative between other invasive and non-invasive hemodynamic monitoring studies}
\begin{tabular}{ccccccc}
\hline 
        Reference & \textup{N.}\ensuremath{^\circ} Patients & Non-Invasive & Rebleeding & Diagnosis required & Real-Time & Mean Precision \\
        Hatib \etal \cite{hatib2018machine} & 1.538 & Yes & No & No & Yes & 0.88 \\
        Kendale \etal \cite{kendale2018supervised} & 13.323 & Yes & No & Yes & N/A & 0.76 \\
        Convertino \etal \cite{convertino2011use}  & 190 & Yes & No & No & Yes & 0.965 \\
        Shoemaker \etal \cite{shoemaker2006outcome} & 661 & Yes & No & Yes & Yes & 0.89 \\
        Prasad \etal \cite{prasad2017prediction} & 101 & No & No & Yes & No & 0.82 \\
        Our work & 548 & Yes & Yes & No & Yes & \textbf{0.98} \\
        \hline 
\end{tabular}
\label{tab:comparative}
\end{table*}

\section{Discussion}

This  study  demonstrated  that  the  use  of  machine learning  models  can  accurately  predict the outcomes  in  acute  stroke  patients and serve to identify the stroke subtype. Many factors influence  stroke outcomes, and these  variables  may have, even a slight, impact  on  prediction. Indeed, based on the data acquired from 798 patients in the Stroke Unit of a large national hospital, this study has demonstrated that machine learning-based models can be derived to help on the right management of this serious neurological disease. 

Our research work has created models for stroke subtype diagnosis, recurrence prediction, and exitus prediction, from hemodynamic data acquired in real time by a traditional monitor. These models can be incorporated into the existing monitors in the Stroke unit to help on neurologist's decisions, or can be used by the emergency services to advance the treatment of the patients. 

This study has derived and evaluated several machine learning-based approaches and, in order to compare and rank the set of algorithms used, we have proposed the hypervolume indicator in the metrics space.

Gradient Boosting algorithm was able to detect hemorrhage vs cerebral infarct within only 30 minutes monitoring, and it achieved an F-Measure of 0.9855. Also, Gradient Boosting algorithm was able to predict a patient’s death in 3 hours with an F-Measure of 0.9988. Finally, Random Forests algorithm obtained an F-Measure of 0.98 to predict a stroke recurrence from 15 minutes of monitoring. 

Table \ref{tab:comparative} compares our study with respect to other similar approaches from the state-of-the-art.  The table shows qualitative and quantitative data, and allows to evaluate the achieved accuracy of the models, if the monitoring approach requires invasive data acquisition, and if the prediction model is able to predict the rebleeding effect in stroke disease. 

As can be seen in the table, our approach not only achieves better accuracy for the developed prediction models, but it is also able to work with non-invasive data and predict the rebleeding events in stroke disease.For  the  first  time  in  literature,  our  real-time  models  from hemodynamic data have opened a new trend in the clinical management of stroke, providing an accurate and efficient approach from the perspective of data processing. 

Our results are a straightforward complement to other approaches in this field that apply ML techniques to predict the stroke event from more complex tests like MRI~\cite{Lee:20}. Our approach, affordable and reliable, could in the future reduce the need of costly neuroimage tools and achieve faster diagnosis.

The present study has several limitations. First, the  patient  population  was not calculated before conduction of the study. However, it should be noted  that  the  current  patient  number has been enough the validate the generalization capabilities of our approach. Second, all the patients had come from one stroke unit. Although these patients may represent the actual characteristics of acute stroke patients, parameters for ML algorithms may have been optimized
and determined for the current population only, suggesting that the results of this study should be interpreted with caution. In a multicenter setting, with different patient populations, further validation is warranted. Finally, patients who arrived at the stroke unit have different evolution time, some of them could have started the stroke episode a few hours ago. This approach should be tested in the initial stages of the event.

\section{Conclusions}
A correct and early stroke type diagnosis is very important in order not to lose the therapeutic window and to correctly send patients to the appropriate medical center.

The prediction of short-term outcomes in ischemic stroke patients may be useful in treatment decisions. In situations where imminent neurological complications could happen, it will allow us to react and change treatment in order to try to avoid it or minimize adverse effects. When death is predicted we can evaluate if there is any therapeutic chance to change it, or if not viable, decide to the adequate therapeutic effort.

In this study, we have also been able to predict the type of stroke with simple semi-intensive routine monitoring analysis using machine learning techniques. No brain image test was evaluated, but it was compared to the doctors' final diagnosis, with very high confidence. It has been also evaluated neurological complications and death risks, with also a very high success rate.

In real clinical practice, it is uncommon to manage acute stroke treatment without a brain image test (CT-scan or MRI). But it is still quite common in some rural areas that early CT-scan access is not possible. This kind of analysis could allow treating before CT-scan performance in case of a high risk of losing the treatment opportunity.

Although in urban areas most patients have easy and fast access to medical attention, depending on the local health organization system, stroke patients not always arrive on time for treatment. Sometimes patients decide to seek direct medical assistance instead of calling 112 (Europe emergency call number), therefore the first medical exam is not performed by a neurologist and a transfer to a stroke unit or stroke center is not possible. So, pre-hospital stroke subtype diagnosis during transfer with routine vital signs monitoring would allow direct transfer to the appropriate medical center in order to receive the necessary treatment. 

%\section*{References}
\bibliography{bibfile}
\pagebreak
\printglossary[title=Abbreviation List,style=mcolindex,type=main,nonumberlist]

\section{Biography}

\begin{wrapfigure}{l}{25mm} 
    \includegraphics[width=1in,height=1.25in,clip,keepaspectratio]{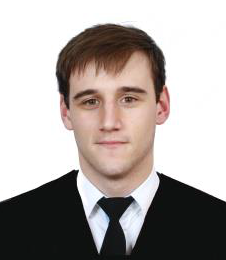}
  \end{wrapfigure}\par
  \textbf{Luis García-Terriza}  received the B.S. degree in Computer Engineering from Complutense University, Madrid, Spain, in 2016 and the M.S. degree in Computer Engineering from  Complutense University, Madrid, Spain, in 2018. He is currently pursuing the Ph.D. degree in Computer Engineering at Complutense University, Computer Science Faculty, Madrid, Spain.
\newline
His research interest includes the development and optimization of Artificial Intelligence models applied to biological/medical area, and parallel and distributed computing. In the private sector, he has been working since 2016 as Data Scientist, Big Data engineer, DevOps Engineer and, lately, Cloud Security Architect.\par

 \begin{wrapfigure}{l}{25mm} 
    \includegraphics[width=1in,height=1.25in,clip,keepaspectratio]{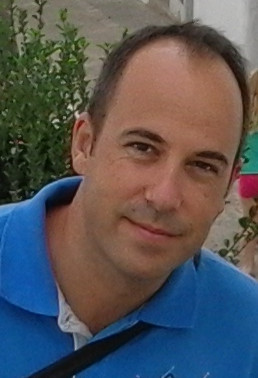}
  \end{wrapfigure}\par
  \textbf{José L. Risco-Martín} is an Associate Professor of the Computer Science Faculty at Universidad Complutense de Madrid, Spain, and head of the Department of Computer Architecture and Automation at the same University. He received his M.Sc. and Ph.D. degrees in Physics from Universidad Complutense de Madrid, Spain in 1998 and 2004, respectively. His research interests focus on Computer Simulation and Optimization, with emphasis on Discrete Event Modeling and Simulation, Parallel and Distributed Simulation, Artificial Intelligence in Modeling and Optimization and Feature Engineering. In these fields, he has co-authored more than 150 publications in prestigious journals and conferences, several book chapters, and three Spanish patents. He has received the SCS Outstanding Service Award in 2017, and the HiPEAC Technology Transfer Award in 2018. He is associate editor of SIMULATION: Trans. of Soc. Mod. and Sim. Int., and has organized several Modeling and Simulation conferences like DS-RT, ANSS, SCSC, SummerSim or SpringSim. He is ACM Member and SCS Senior Member. Prof. José Luis Risco Martín has participated in more than 15 research projects and more than 10 contracts with industry. He has elaborated simulation and optimization models for companies like Airbus, Repsol or ENAGAS. \par
  
   \begin{wrapfigure}{l}{25mm} 
    \includegraphics[width=1in,height=1.25in,clip,keepaspectratio]{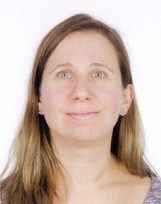}
  \end{wrapfigure}\par
  \textbf{Gemma Reig Roselló} received the MD degree in Medicine and Surgery from University of Barcelona (UB), Spain, in 2002. Becoming a Neurologist in 2007, after Neurology residency program in Hospital de La Princesa, in Madrid, Spain. Became Stroke Fellow for two years after residency in Hospital de La Princesa. Being able to join as a staff member in the Neurology department, working as a Stroke neurologist since 2009. She is a  clinical teaching collaborator in the Neurology Department, teaching students and residents, in theory classes and clinical practices. She is also an Associate Member of Asociación Madrileña de Neurología – Madrid Neurology Association (AMN) and Sociedad Española de Neurología- Spanish Neurology Society (SEN). She has participated in a large number of international and national research projects and pharmaceutical clinical trials, mainly in stroke field. Her research interests include all field of neurology, but specially stroke, and also from a computer and engineering point of view, having the chance to work with big data tools and non-invasive monitoring studies. She is active in publishing posters and oral communications in Neurology medical conferences. \par

   \begin{wrapfigure}{l}{25mm} 
    \includegraphics[width=1in,height=1.25in,clip,keepaspectratio]{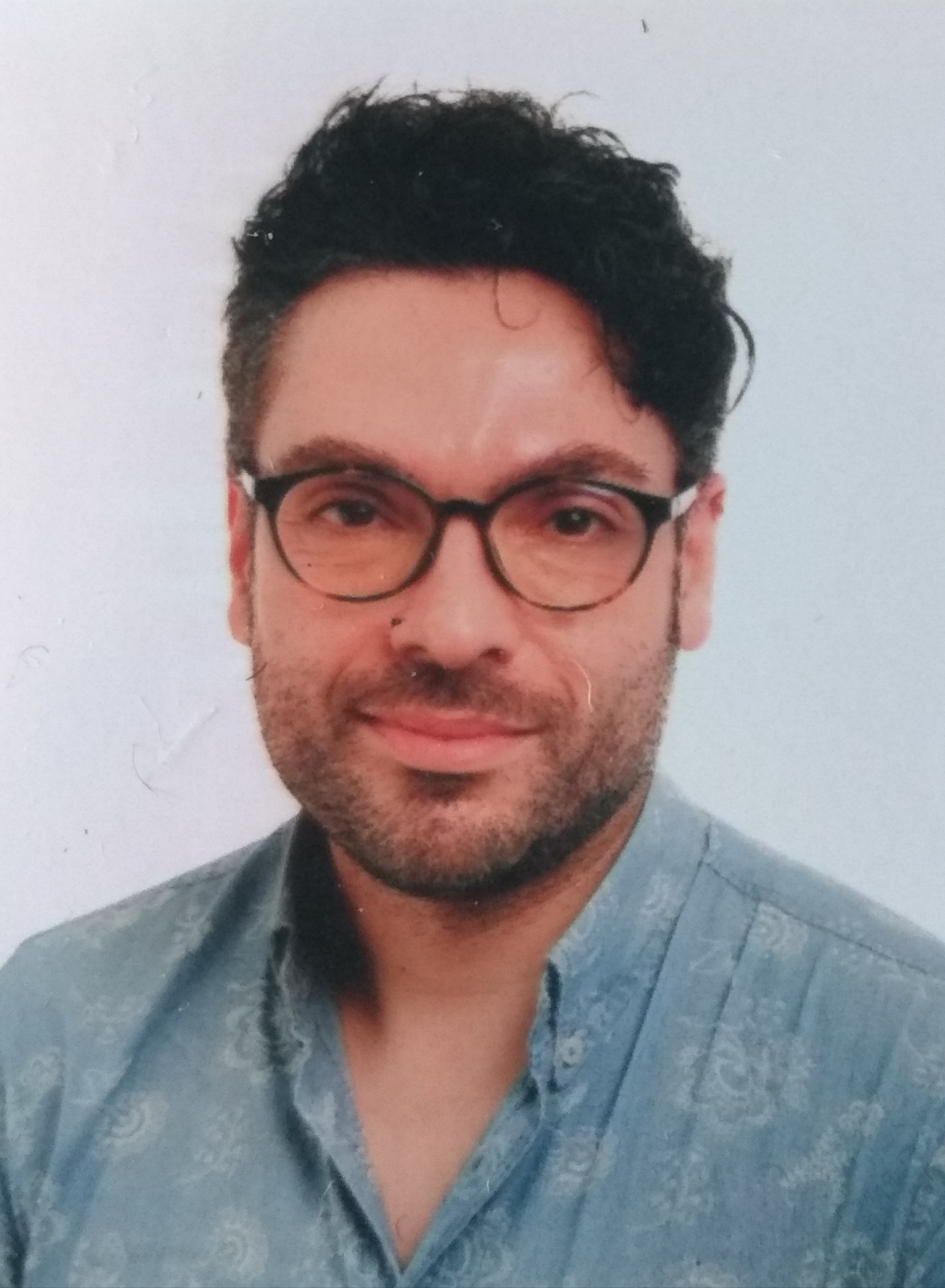}
  \end{wrapfigure}\par
  \textbf{Jose L. Ayala} received the M.Sc. and Ph.D.degrees in telecommunication engineering from the Technical University of Madrid (UPM), Spain, in 2001 and 2005, respectively. His PhD dissertation was awarded by the Council of Telecommunication Engineers, and the ACM Special Interest Group on Design Automation. He is currently an Associate Professor in the Department of Computer Architecture and Automation, Universidad Complutense de Madrid, Madrid, Spain. He is also an Associate Member of the Center for Computational Simulation (UPM), and the IEEE Council of Electronic Design Automation. He also participates actively in the IEEE IoT, Smart Cities, and Brain initiatives. He has served as organizing committee of many international conferences, including DATE, GLSVLSI, VLSI-SoC, ISC2, ISLPED. He has led a large number of international research projects and bilateral projects with industry, in the fields of power and energy optimization of embedded systems, and noninvasive health monitoring. His research interests include automatic diagnostic tools, predictive modeling in the biomedical field, and wearable non-invasive monitoring. Prof. Ayala has received 4 best paper awards, and more than 15 awards related to technology transfer. He is also CEO of the high-tech start-up, BrainGuard, devoted to the development of predictive models from ambulatory data in the neurological field. \par

\end{document}